# Spectroscopic Signature for Local-moment Magnetism in van der Waals Ferromagnet Fe$_3$GeTe$_2$

X. Xu[1*], Y. W. Li[2*], S. R. Duan[1], S. L. Zhang[3], Y. J. Chen[1], L. Kang[1], A. J. Liang[3,4], C. Chen[3,4], W. Xia[3], Y. Xu[1,5,6], P. Malinowski[7], X. D. Xu[7,8], J.-H. Chu[7], G. Li[3], Y. F. Guo[3], Z. K. Liu[3], L. X. Yang[1,5*], and Y. L. Chen[1,2,4†]

[1]State Key Laboratory of Low Dimensional Quantum Physics, Department of Physics, Tsinghua University, Beijing 100084, P. R. China
[2]Department of Physics, Clarendon Laboratory, University of Oxford
Parks Road, Oxford, OX1 3PU, UK
[3] School of Physical Science and Technology, ShanghaiTech University and CAS-Shanghai Science Research Center, Shanghai 201210, China
[4]Advanced Light Source, Lawrence Berkeley National Laboratory, Berkeley, California 94720, USA
[5]Collaborative Innovation Center of Quantum Matter, Beijing 100084, P. R. China
[6]RIKEN Center for Emergent Matter Science (CEMS), Wako, Saitama 351-0198, Japan
[7]Department of Physics, University of Washington, Seattle, WA 98105, USA
[8]Department of Materials Science and Engineering, University of Washington, Seattle, WA 98105, USA

*lxyang@tsinghua.edu.cn          †yulin.chen@physics.ox.ac.uk

**The van der Waals ferromagnet Fe$_3$GeTe$_2$ has recently attracted extensive research attention due to its intertwined magnetic, electronic and topological properties. Here, using high-resolution angle-resolved photoemission spectroscopy, we systematically investigate the temperature evolution of the electronic structure of bulk Fe$_3$GeTe$_2$. We observe largely dispersive energy bands that are narrowed by a factor of 1.6 compared with ab-initio calculation. Upon heating towards the ferromagnetic transition near 225 K, we observe a massive reduction of quasiparticle coherence in a large energy range, which is attributed to the enhanced magnetic fluctuation in the system. Remarkably, the electron bands barely shift with increasing temperature, which deviates from exchange splitting picture within the itinerant Stoner model. We argue that the local magnetic moments play a crucial role in the ferromagnetism of Fe$_3$GeTe$_2$, despite its strongly itinerant nature. Our results provide important insights into the electronic and magnetic properties of Fe$_3$GeTe$_2$ and shed light on the generic understanding of itinerant magnetism in correlated materials.**

The magnetism not only is a fascinating quantum phenomenon in itself, but also immensely influences various emergent properties such as unconventional superconductivity [1-3], heavy fermion systems [4, 5], topological quantum physics [6, 7] and quantum critical behaviours[8, 9]. In order to properly describe the magnetism in electronic materials, two paradigmatic frameworks have been established, concentrating on two opposing extremes: itinerant and local magnetism [10]. Within the weak-coupling itinerant picture as represented by well-known Stoner model, the spin-polarized exchange splitting of electron bands drives the long-range magnetic ordering in metallic systems, while the local magnetic moments take charge of the magnetism mainly in insulating materials according to the localized Heisenberg model. However, to distinguish these two mechanisms is usually challenging, especially in correlated materials, where the localized magnetic moments, although screened by itinerant electrons, strongly modify the quasiparticle energy bands [11]. Such competition between local and itinerant magnetism is well demonstrated by the longstanding debate regarding the nature of magnetic ordering in the cuprate and iron-based superconductors [3, 12-15].

Recently, atomically thin van der Waals ferromagnets have attracted great research attention because of their survival against thermal fluctuation, which provides a versatile flatland to explore the nature of the magnetism that is decorated by different correlation strength and promises a rich material basis for future application in electronic and spintronic devices. Up to date, many different types of van der Waals ferromagnetic materials, such as $Cr_2Si_2Te_6$/$Cr_2Ge_2Te_6$, $CrI_3$/$CrBr_3$, and $Fe_3GeTe_2$ (FGT), have been thinned to their physical limit to realize two-dimensional ferromagnets [16-24]. Among them, FGT has been intensively studied due to the realization of tuneable room temperature ferromagnetism in its thin films [21, 22]. Besides, bulk FGT also exhibits fertile and intriguing properties, such as extremely large anomalous Hall effect induced by topological nodal lines [25], Kondo lattice physics [26], strongly enhanced electron mass [27], and magnetocaloric effect [28]. Although it is widely believed that the ferromagnetism in FGT is itinerant in nature

[29], a compelling consensus regarding the mechanism of the ferromagnetic ordering is yet not established, as the ferromagnetism in FGT can be understood within a classical Heisenberg model [21]. There is even debate regarding whether Fe atoms align ferromagnetically or antiferromagnetically in FGT [30]. Since the electronic structure can provide crucial insights into magnetic properties, it is elucidative to investigate the electronic structure of magnetic materials and its temperature evolution, which is still not adequate enough in FGT.

In this Letter, we systematically investigate the electronic structure of bulk FGT and its temperature evolution using high-resolution angle-resolved photoemission spectroscopy (ARPES) and *ab-initio* band structure calculation. The comparison between experiment and density functional theory (DFT) calculation suggests a moderate electronic correlation in FGT. The scattering rate in the system quickly intensifies and the quasiparticle spectral weight is strongly suppressed upon heating towards the Curie temperature ($T_C$) near 225 K, which are attributed to the enhancement of magnetic fluctuation in the system. In contrast to the temperature dependent exchange splitting within the Stoner model, we observe minor change of the shape and position of band dispersions with temperature. Therefore, we argue that the local magnetic moments and strong spin fluctuation are crucial in the ferromagnetism of FGT, which provides important insights into the nature of ferromagnetism and other novel properties of FGT.

High-quality FGT single crystals with size of $5 \times 5 \times 0.1$ mm$^3$ were synthesized by chemical transport method with iodine as transport agent [22]. ARPES data were taken with various photon energies at beam line I05 of Diamond Light Source (DLS) under proposal No. SI20683-1, beamline 9A of Hiroshima Synchrotron Radiation Center (HSRC) and beamline 5-2 of Stanford Synchrotron Radiation Laboratory (SSRL). Scienta R4000 electron analysers are equipped at all three beamlines. The overall energy and angular resolutions are 15 meV and 0.3°. Electronic structures were calculated using DFT as implemented in the Vienna ab initio simulation package (see the supplementary file for details).

FGT crystallizes into a layered hexagonal structure with the space group of *P63/mmc* (No. 194). There are two inequivalent Fe atoms in the unit cell of FGT. In the ferromagnetic state, the magnetic moment of all the Fe atoms aligns along the *c* axis below 225 K [28-30]. Figure 1 shows the magnetic transport measurement on FGT with field applied along the *c* axis. We observe clear magnetic hysteresis loop showing the ferromagnetic nature of FGT. The magnetic moment saturates to a value of about 1.6 $\mu_B$/Fe above 0.4 T (Fig. 1(a)). Figure 1(b) shows the temperature dependence of the magnetization. The $T_c$ of FGT is determined to be about 225 K, at which the temperature dependent resistivity shows an anomaly (Fig. 1(c)), in consistence with previous measurements [27-29]. Below $T_c$, FGT exhibits a large anomalous Hall effect, as shown by the temperature dependent anomalous Hall angle in Fig. 1(d), which has been attributed to the topological nodal lines in the system [25].

Figure 2 investigates the electronic structure of FGT in the ferromagnetic state at 10 K. Despite the weak interlayer coupling, we observe strong $k_z$ variation of the energy bands in Fig. 2(a) (see Supplemental Material [31]). Figure 2(b) shows the Fermi surface in the $k_x$-$k_y$ plane measured with 114 eV photons that enhance the states near $\bar{K}$. We observe a hexagonal Fermi pocket near the $\bar{\Gamma}$ point, with complex texture structure inside. There is a small electron pocket Near the $\bar{K}$ point and a distribution of blurred spectral weight near the $\bar{M}$ point, in consistence with previous ARPES measurement and band structure calculation [26]. Figures 2(c)-(e) show the band dispersions along high symmetry directions. Due to the multi-orbital nature and $k_z$ dispersion of the energy bands, the measured Fermi surface and band dispersions strongly depend on the photon polarization and photon energy (see the supplementary file). Using 114 eV photons with linearly vertical polarization, we observe mainly four bands in an energy range of 1.2 eV below the Fermi energy ($E_F$). The α and γ bands contribute to the hole pockets near $\bar{\Gamma}$, while the δ band contributes to the small electron pocket near $\bar{K}$. Our DFT calculation shows the complex band structure of FGT in the ferromagnetic state (Fig. 2(f)). We note that the calculated bands need to be renormalized by a factor of about 1.6 in order to obtain an overall agreement with the experiment, indicating a moderate correlation effect in FGT (see Supplemental Material [31]). After renormalization, our calculation is consistent with previous dynamical mean-field theory (DMFT) calculations that suggest a relatively large Hubbard interaction in Fe sites and confirm the importance of electron correlation in the ferromagnetism of FGT [25, 27]. By comparing the calculated and measured δ band, we obtain an electron effective mass enhancement by a factor of $1.65 \pm 0.04$ (see Supplemental Material [31]), and we do not observe mass enhancement more significant than this value from the Fermi velocity of other bands. This value is much smaller than that obtained from Sommerfeld coefficient, which keeps mysterious [27, 29]. The calculated band structure in the ferromagnetic state shows a large exchange splitting (~ 1.5 eV as indicated by the arrows in Fig. 2(f)) compared with the nonmagnetic state (Fig. 2(g)), which is believed to drive the ferromagnetic ordering in FGT according to Stoner mechanism [29].

In order to unveil the nature of the ferromagnetism in FGT, we track the temperature evolution of its electronic structure in Fig. 3. Figures 3(a) and 3(b) show the band dispersions around $\bar{\Gamma}$ and $\bar{K}$ at selected temperatures measured with 27 eV (horizontally polarized) and 114 eV (vertically polarized) photons, respectively. We observe different bands η and β near Γ due to the $k_z$ dispersion and polarization dependence of energy bands (see Supplemental Material [31]). Both the η and δ bands slightly shift towards lower binding energies with spectral weight strongly suppressed at high temperatures. Figures 3(c) and 3(d) show the temperature evolution of the energy distribution curves (EDCs) near the $\bar{\Gamma}$ and $\bar{K}$ points, respectively. The EDC peak intensity quickly decreases with increased temperature, as presented in Fig. 3(e). In addition, the ARPES spectra are strongly blurred upon warming towards $T_C$, suggesting a dramatic enhancement of the disordering level in the system, as estimated by the quickly increased momentum distribution curve (MDC) width in Fig. 3(f). Both the suppression of the quasiparticle spectral weight and the increase of MDC width show an anomaly near $T_C$, suggesting an intimate correlation between the broadening of the spectra with the magnetism in the system. Thus, we attribute the observed spectra broadening into the enhancement of the magnetic fluctuation in the system with increased temperature [29]. Notably, we observe the broadening of the energy bands in a large energy range, which is out of the expectation of itinerant spin fluctuations that mainly affect the states near $E_F$.

Within the Stoner model, a prototypical itinerant ferromagnet is expected to exhibit a temperature dependent exchange splitting of the energy bands (Figs. 2(f) and 2(g)). However, we do not observe considerable change of the electronic structure with temperature in Fig. 3, except for band broadening and minor band shift, as further tracked by MDCs and EDCs in Figs. 4(a) and 4(b). Despite the strong broadening of the bands at high temperatures, we can still resolve the band dispersions even above $T_C$. We summarize the band shift extracted from MDCs and EDCs in Fig. 4(c) to compare with the temperature dependent exchange splitting estimated from the magnetic moment that is scaled to half of the DFT calculated exchange splitting (about 1.5 eV). We observe a minuscule band shift that strongly deviates from the expected temperature evolution of the exchange splitting within the Stoner model (Fig. 4(c), see Supplementary Information for temperature evolution of more bands). This result is in contrast to the canonical itinerant ferromagnets such as Fe and Ni, where enormous temperature dependent band shift has been revealed [32-35]. On the contrary, our observation fits better to the temperature independent model based on the localized exchange interaction as indicated by the horizontal dashed line in Fig. 4(c) [32]. Clearly, a completely itinerant magnetic mechanism is not capable to reproduce our experiment. The local magnetic moment, on the other hand, should be taken into account in order to properly describe the ferromagnetism in FGT. Our conclusion is supported by the observation of Kondo lattice behaviour and the explanation of the ferromagnetism in FGT within the Heisenberg model [21, 26].

Our results mimic the behaviours of other itinerant ferromagnets such as $SrRuO_3$, in which strong mass enhancement, electron decoherence and local magnetism indicated by minor band shift with temperature are observed [36]. However, the underlying microscopic interaction and the impact of local moments on the electronic structure are different in these two systems. The quasiparticle effective mass is strongly renormalized by electron-boson interaction in SrRuO$_3$, as demonstrated by a kink structure in the band dispersion. On the contrary, we did not observe any kink structure in FGT and the electron effective mass is only enhanced by a factor of 1.65 due to electronic correlation effect, although the specific heat measurement alludes a dramatic enhancement of quasiparticle effective mass [27, 29]. It is the electron-electron interaction that is more important in the magnetism of FGT [27], since the effective on-site Coulomb interaction between Fe 3$d$ electrons is much stronger than Ru 4$d$ electrons.

Moreover, we observe an acceleration of the temperature dependent band shift near $T_c$, which can be due to the collapse of long range ferromagnetic ordering due to the enhanced magnetic fluctuation near $T_c$. Yet, the band shift is still insignificant compared with that estimated from magnetization (about 10%), suggesting an important influence of the local moments on the electronic structure of FGT even without long-range ordering. This observation is similar to the situation in Fe and Ni when the long-range order collapses due to the magnetic fluctuation near $T_c$, where the temperature dependent band shift ceases due to the revival of local moments [36].

In conclusion, we have presented the temperature evolution of the electronic structure of a van der Waals ferromagnet FGT. We observe strong electron mass enhancement, substantial electron decoherence in a large energy range and minor band shift with temperature, which are against a weak-coupling itinerant picture. We argue that the local magnetic moments should play important roles in the itinerant magnetic systems, which will deepen our generic understanding of the magnetism in condensed materials.

This work was supported by the National Natural Science Foundation of China (Grant No. 11774190, No. 11674229 and No. 11634009), the National Key R&D program of China (Grant No.

2017YFA0304600 and 2017YFA0305400), EPSRC Platform Grant (Grant No. EP/M020517/1). The work at University of Washington was supported by NSF MRSEC at UW (DMR-1719797) and the Gordon and Betty Moore Foundation's EPiQS Initiative, Grant GBMF6759 to J.-H.C. Y.F. G. acknowledges the support from the Shanghai Pujiang Program (Grant No. 17PJ1406200). L. X. Y. acknowledges the support from Tsinghua University Initiative Scientific Research Program.

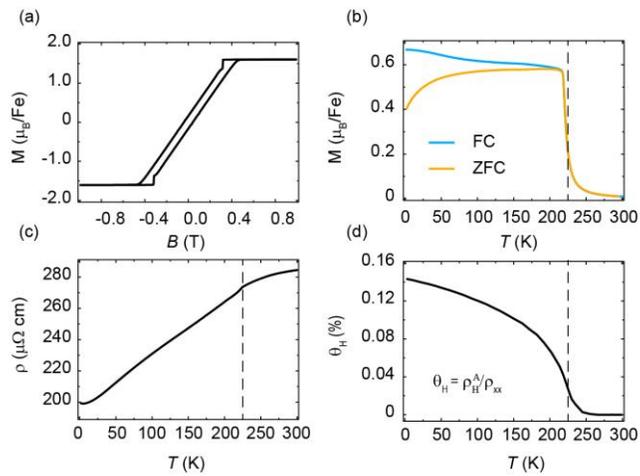

FIG. 1 (Color online). (a) Magnetic moment as a function of magnetic field for FGT measured at 2 K showing the magnetic hysteresis loop. (b) Temperature dependent magnetization of FGT measured under zero-field-cooling and field-cooling conditions. (c) Temperature dependent resistivity of FGT. (d) Temperature dependent anomalous Hall angle of FGT. All the data were collected with the magnetic field applied along the c axis. The vertical dashed line indicates the Curie temperature.

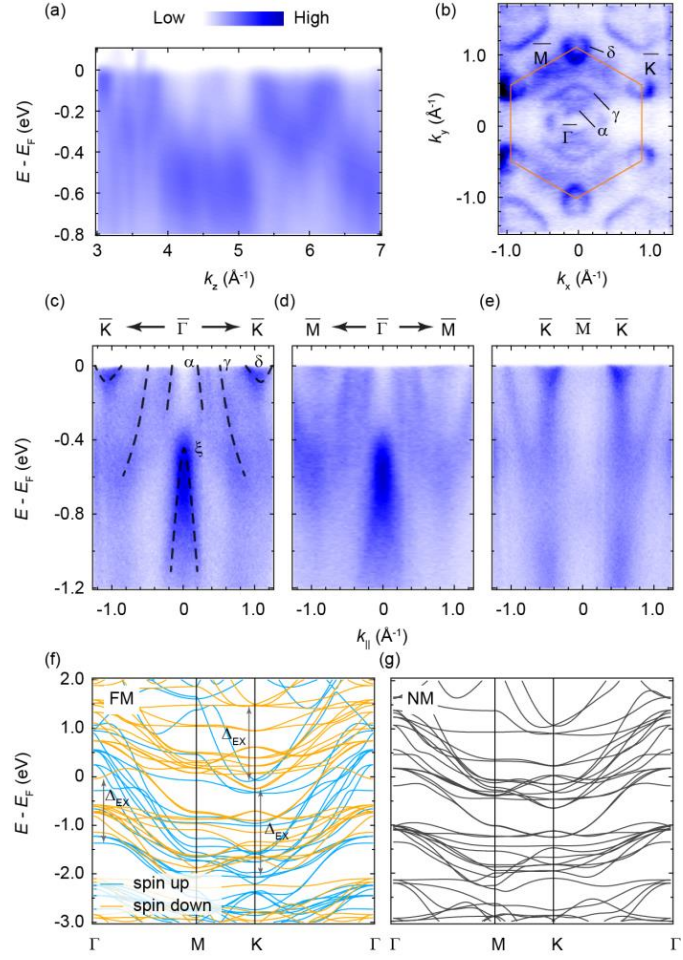

FIG. 2 (Color online). (a) Band structure along $\Gamma A$ showing strong $k_z$ dispersion in FGT. (b) Fermi surface map obtained by integrating ARPES intensity in an energy window of 20 meV near the Fermi energy ($E_F$). (c)-(e) Band dispersion along different high symmetry directions. The black dashed curves are guides to the eyes for the band dispersions. (f), (g) DFT calculation of the band dispersions in ferromagnetic (FM) and nonmagnetic (NM) states. Data in panels (a) and (b)-(e) were collected using 114 eV photons with linearly horizontal (a) and vertical (b-e) polarizations.

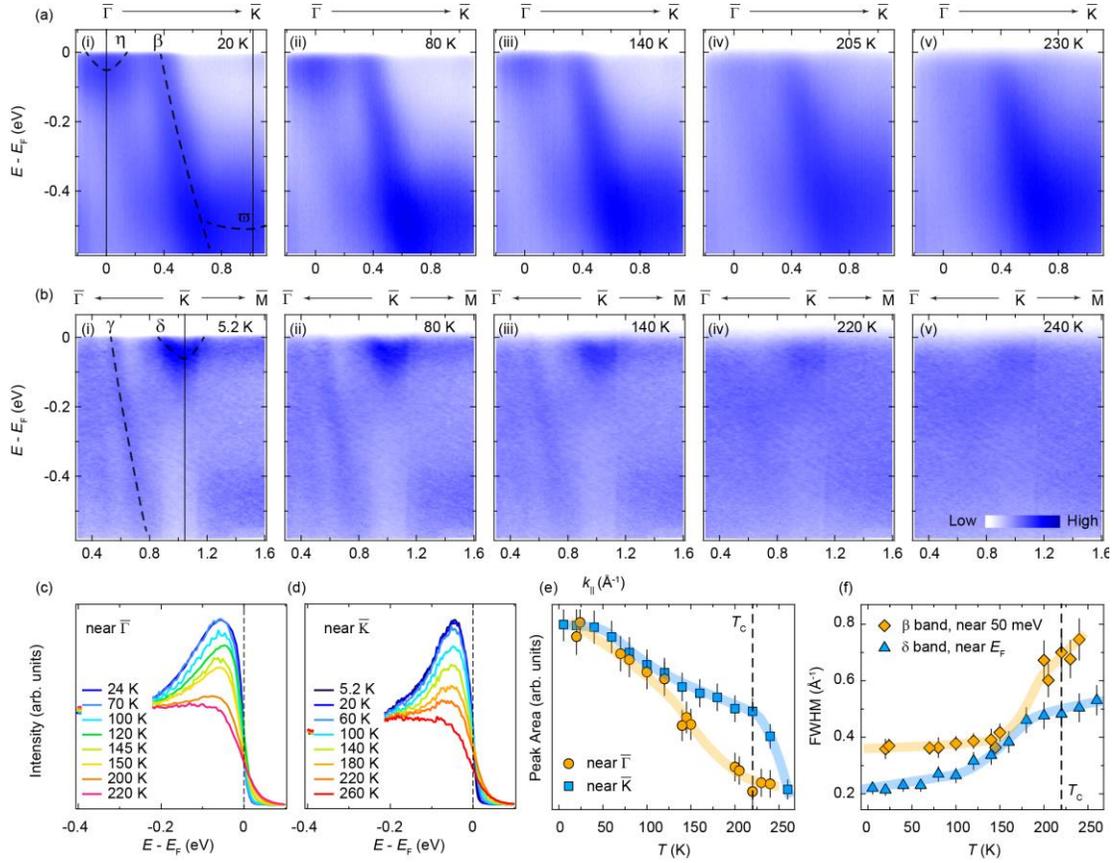

FIG. 3 (Color online). (a), (b) Temperature evolution of the band dispersions near the $\overline{K}$ (a) and $\overline{\Gamma}$ (b) points. (c), (d) Energy distribution curves (EDCs) integrated in a momentum window of 0.25 Å$^{-1}$ near the K (c) and Γ (d) points. (e) The peak area of the EDCs in panels (c) and (d) as a function of temperature. EDC peak area is obtained by integrating the EDC in an energy window of 200 meV. (f) The full width at half maximum (FWHM) of the momentum distribution curves (MDCs) as a function of temperature. The MDCs are integrated in an energy window of 45 meV near $E_F$ and fitted with a Lorentzian to extract its FWHM. The orange and blue curves are the guide to eyes for the temperature evolution of peak area and MDC width. Data in panels (a) and (b) were collected with photons at 27 eV and 114 eV respectively.

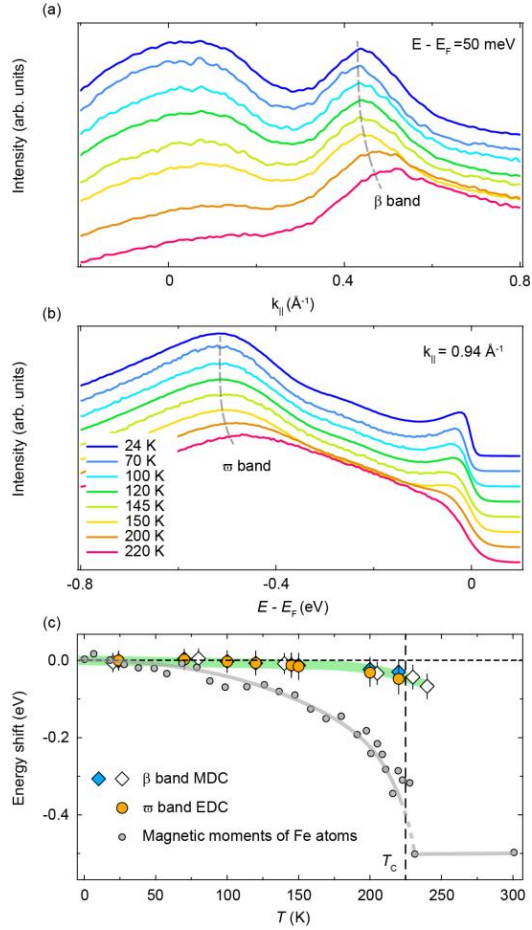

FIG. 4 (Color online). (a) MDCs integrated in an energy window of 50 meV near -50 meV at selected temperatures. (b) EDCs integrated in a momentum window of 0.15 Å⁻¹ near 0.94 Å⁻¹ at selected temperatures. The gray dashed curves in panel (a) and (b) are guides to eyes for the shift of EDC and MDC peaks with temperature. (c) The shift of energy bands as extracted form MDCs and EDCs as a function of temperature. The gray circles are the magnetic moemnts in FGT measured with neutron scattering as reproduced from Ref. 28. The gray and green curves are guides to eyes for the temperature dependence of the magnetization and measured band shift. The temperature independent dashed line shows the expected temperature evolution of the energy shift within a completely localized ferromagnet.